\documentstyle[psfig,conf_iap]{article}
\begin{document}

\def\BE{\begin{equation}}
\def\BEL#1{\begin{equation}\label{#1}}
\def\EE{\end{equation}}
\newcommand{\BAD}[1]{\emph{#1}}
\newcommand{\MAG}{{\rm ~mag}}
\newcommand{\HI}{H\,{\scriptsize I}}
\newcommand{\etal}{et al.}
\newcommand{\arcmin}{'}
\newcommand{\degree}{^\circ}
\newcommand{\K}{{\rm ~K}}
\newcommand{\cm}{{\rm ~cm}}
\newcommand{\micron}{~\mu {\rm m}}
\newcommand{\nWpMMSr}{{\rm ~nW}~{\rm m}^{-2}~{\rm sr}^{-1}}
\newcommand{\MJy}{{\rm ~MJy}}
\newcommand{\MJypSr}{{\rm ~MJy}~{\rm sr}^{-1}}
\newcommand{\Jy}{{\rm ~Jy}}
\newcommand{\BminusV}{B{\rm -}V}
\newcommand{\Mgtwo}{{\rm Mg}_2}
\newcommand{\Xmap}{{\bf X}} 
\newcommand{\Icorr}{{\bf I}_{100}} 
\newcommand{\Ebv}{E(\BminusV)}
\newcommand{\Ab}{A(B)}

\heading{%
%
Application of SFD Dust Maps to Galaxy Counts and CMB Experiments
%
} 
\par\medskip\noindent
\author{%
David Schlegel$^{1,2}$, Douglas Finkbeiner$^{3}$, Marc Davis$^{3}$
}
\address{%
Princeton University, Dept.\ of Astrophysical Sciences,
Princeton, NJ 08544 USA
}
\address{%
University of Durham, Dept.\ of Physics, South Road, Durham DH1 3LE UK
}
\address{%
University of California, Dept.\ of Astronomy, Berkeley, CA 94720 USA
}

\begin{abstract}
We have constructed a full-sky map of the far-infrared suitable for
measuring Galactic reddening and extinction (\cite{sfd98}: SFD).
The SFD map is based upon extensive re-analysis of data from the
COBE/DIRBE and IRAS satellite missions.
We demonstrate that the maps can correct for extinction problems in
the APM galaxy survey.  We also determine the most dust-free regions
for conducting cosmic microwave background or soft X-ray experiments
of extragalactic objects.

\end{abstract}

\section{Introduction}

Dust in the Milky Way is a pernicious contaminant to many extra-galactic
observations.
These dust grains absorb and scatter UV and optical light, both extincting
and reddening extragalactic objects.  At microwave frequencies, the dust
emits radiation that contaminates studies of the cosmic microwave background
(CMB).

The different components of radiation in the night sky  from near-infrared through
microwaves are shown in Figure \ref{fig_cib}.
The Galactic dust spectrum is significantly non-thermal at wavelengths
$\lambda < 100 \micron$ due to stochastic heating of very small grains
(VGSs) \cite{draine85}.  At wavelengths $100 < \lambda < 300 \micron$,
the dust spectrum is well-fit by a thermal spectrum.
The typical color temperature of this dust is $18.2\K$ for an $\alpha=2$
emissivity law.

We have fit the Galactic dust temperature and column depth using
the emission at $100$ and $240\micron$ as measured by DIRBE.
The dust emission marginally exceeds all other sources of emission
at these wavelengths.

\begin{figure}
\centerline{\vbox{
\psfig{figure=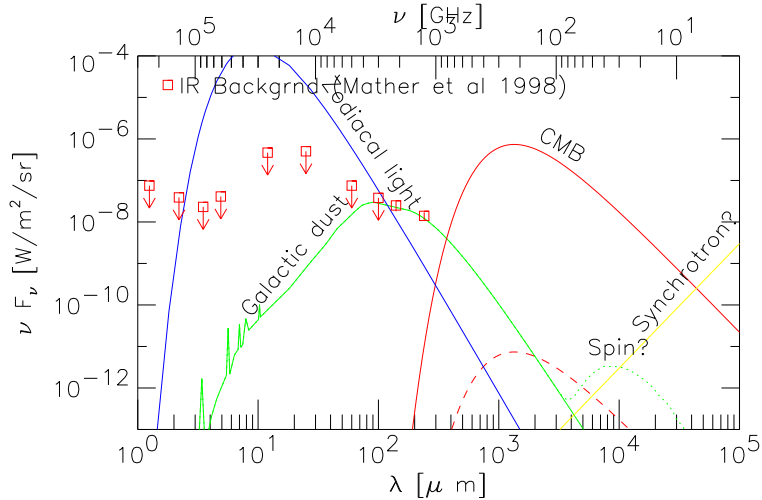,width=10.0cm}
}}
\caption[]{Contributions to thhe infrared sky above the Earth's atmosphere.
}
\label{fig_cib}
\end{figure}

\section{Generating the Dust Map}

Our Galactic dust map is based upon extensive re-analysis of data from the
COBE/DIRBE \cite{boggess92} and IRAS \cite{neugebauer84} satellite missions.
Accurately determining the column of the diffuse Galactic dust emission would
not have been possible before the COBE satellite flew in 1989.
The COBE/DIRBE instrument mapped the entire sky from $2$ to $240\micron$
with excellent controls of artifacts, zero-points, and calibrations.
The older IRAS data does have the one benefit of superior angular resolution
($\sim 4\arcmin$ as compared to $0.{\degree}7$) for $96\%$ of the sky
at $100\micron$.  We combine the DIRBE and IRAS maps in such a manner
as to achieve a resolution of $6.{\arcmin}1$ while maintaining the DIRBE
zero-point and calibration.

The major difficulty is decoupling the dust emission from zodiacal light
and the cosmic IR background (see Figure \ref{fig_cib}).
Details of our methods can be found in Schlegel, Finkbeiner \& Davis
(\cite{sfd98}: SFD).
The maps and associated software have been made freely available at \\
{\tt http://astro.berkeley.edu/davis/dust/index.html} \\
{\tt http://astro.princeton.edu/{$\sim$}schlegel/dust/index.html}

\section{Measuring Reddening}

\subsection{Calibrating the Map}

We have computed the column density of dust as the amount of $100\micron$
emission expected if all the dust were at the same reference temperature
($T_0 = 18.2\K$).  We then normalize the amplitude of reddening
per unit of $100\micron$ flux.
Our reddening estimates can be written as
\BE \Ebv = p~\Icorr \Xmap , \EE
where $\Icorr$ is the point source-subtracted IRAS-resolution
$100\micron$ map, $\Xmap$ is the temperature-correction map,
and we seek the calibration coefficient $p$.
The reddening of external galaxies allows the most straightforward
calibration of $p$.  For a sample of 389 galaxies provided by Dave Burstein,
we have measured $p = 0.0184 \pm 0.0014 \MAG / \MJypSr$.

\subsection{Correcting APM Magnitudes}

The APM galaxy survey \cite{maddox90b} has measured
photometric properties for 3 million galaxies.  The original survey region
($\delta < -20\degree$, $|b| > 40\degree$) was chosen to be in a very clean
part of the sky.  We have shown previously (\cite{schlegel95}; \cite{finkbeiner97})
that the scientific conclusions from this region,
such as the galaxy-galaxy correlation function, are not
affected by Galactic extinction.
However, the full survey includes 84 plates at lower latitudes
($30\degree < |b| < 40\degree$) that are subject to extinction problems.

Figure \ref{fig_apm18} plots the APM galaxy counts (dotted line)
versus the reddening as predicted by SFD in a magnitude slice
$17.5 < b_J < 18.5$.  The galaxy counts drop off with reddening as
one would predict (solid line).  Given the positions and magnitudes
for each of the 3 million galaxies, we have attempted to correct the
measured magnitudes individually for each galaxy.
If we then re-select
galaxies in the same magnitude slice, the correlation with reddening
has been almost entirely removed (dashed line).  Attempting such
photometric corrections with the BH maps does not remove the correlation
(dot-dash line).

\begin{figure}
\centerline{\vbox{
\psfig{figure=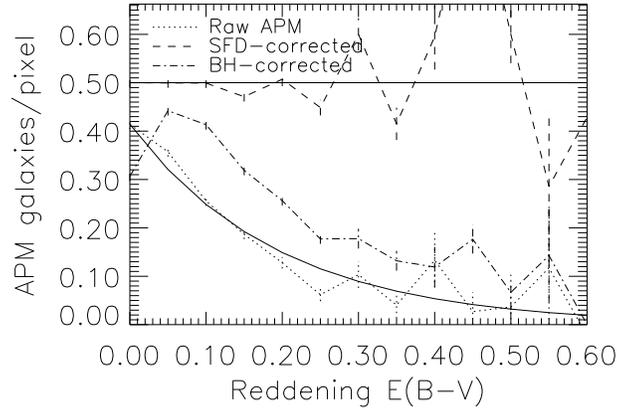,width=9.0cm}
}}
\caption[]{APM counts versus predicted reddening.
}
\label{fig_apm18}
\end{figure}

\section{Dust-Free Regions?}

We find most of the sky to have a measurable column density of dust.
At high Galactic latitudes ($|b| > 45 \degree$), we find far-IR emission
corresponding to a median reddening of $\Ebv = 0.020 \MAG$ or an extinction
of $\Ab = 0.086\MAG$.
This disagrees with the claim by Burstein \& Heiles \cite{bh82} that
these high-latitude zones have zero reddening.

We do find regions in the Southern hemisphere that have dust columns
four times lower than the high-latitude average and two times lower
than the Lockman hole.  These regions, near $l=240\degree$, $b=-49\degree$
(see Table \ref{table_holes})
should be prized zones for CMB or soft X-ray measurements.
Note that 21-cm data does \emph{not} exist to confirm that these zones
are also \HI\ minima.

The north celestial pole is a particularly poor region for conducting
CMB experiments.  Across this pole extends the Polaris Flare, which
is both dusty and $2\K$ colder than the surrounding medium.  Because
it is colder, we expect its mm-wave emission to be larger than one would
estimate from its $100\micron$ flux.

\begin{table}
\footnotesize
\begin{center}
\begin{tabular}{|r r r r r r l|}
\hline
   {$\alpha_{2000}$}     &
   {$\delta_{2000}$}     &
   {$l$}                 &
   {$b$}                 &
   {$\Icorr \Xmap$}      &
   {$N($\HI$)$}          &
   {Comments}            \\
   {(hr)}                &
   {(deg)}               &
   {(deg)}               &
   {(deg)}               &
   {($\MJypSr$)}         &
   {($10^{19} \cm^{-2}$)} &
   {}                    \\
\hline
 0 28 & -42 44 & 318.4 & -73.7 &  0.35 &  ... &                 \\
 0 51 & -27 08 &   0.0 & -90.0 &  0.79 & 15.5 &  SG pole        \\
 3 57 & -48 50 & 257.1 & -48.4 &  0.33 &  ... &                 \\
 3 59 & -42 47 & 248.0 & -49.2 &  0.30 &  ... &                 \\
\BAD{ 4 01} & \BAD{-34 25} & \BAD{235.2} & \BAD{-49.1} &  \BAD{0.20} &
  \BAD{...} &                 \\
\BAD{ 4 03} & \BAD{-37 37} & \BAD{240.0} & \BAD{-48.6} &  \BAD{0.17} &
  \BAD{...} &                 \\
\BAD{ 4 05} & \BAD{-35 50} & \BAD{237.4} & \BAD{-48.2} &  \BAD{0.20} &
  \BAD{...} &                 \\
\BAD{ 4 44} & \BAD{-53 20} & \BAD{261.3} & \BAD{-40.2} &  \BAD{0.23} &
  \BAD{...} &                 \\
10 36 & +56 38 & 152.7 &  52.0 &  0.29 &  4.4 &                 \\
10 48 & +57 02 & 150.5 &  53.0 &  0.44 &  5.8 &  Lockman hole   \\
12 51 & +27 08 &   0.0 &  90.0 &  0.66 & 10.1 &  NG pole        \\
13 35 & +39 09 &  88.4 &  74.9 &  0.27 &  8.6 &                 \\
13 42 & +40 30 &  88.0 &  73.0 &  0.27 &  8.5 &                 \\
13 44 & +57 04 & 109.2 &  58.6 &  0.28 & 10.3 &                 \\
13 54 & +41 33 &  85.2 &  70.6 &  0.29 &  9.1 &                 \\
14 10 & +39 33 &  75.3 &  69.5 &  0.29 &  7.2 &                 \\
22 43 & -46 49 & 346.3 & -58.1 &  0.41 &  ... &                 \\
23 22 & -46 22 & 339.4 & -64.0 &  0.38 &  ... &                 \\
\hline
\end{tabular}
\caption{Regions of low dust column density (averaged in $1\degree$
aperatures).}
\label{table_holes}
\end{center}
\end{table}

\vfill
\end{document}